\documentclass{ws-procs975x65}

\def\beq{\begin{equation}}
\def\eeq{\end{equation}}
\def\xx{\mathrm{x}}
\def\yy{\mathrm{y}}

\begin{document}

\title{Observables from modified dispersion relations on curved spacetimes: circular orbits, redshift and lateshift}

\author{Christian Pfeifer} 

\address{Laboratory of Theoretical Physics, Institute of Physics, University of Tartu, W. Ostwaldi 1\\
50411 Tartu, Estonia\\
E-mail: christian.pfeifer@ut.ee}

\begin{abstract}
The Hamiltonian formulation of modified dispersion relations (MDRs) allows for their implementation on generic curved spacetimes. In turn it is possible to derive phenomenological effects. I will present how to construct the kappa-Poincare dispersion relation on curved spacetimes, its spherically symmetric realizations, among them the kappa deformation of Schwarzschild spacetime, and its implementation on Friedmann-Lemaitre-Robertson-Walker spacetimes with arbitrary scale factor. In addition we will construct the general first order modifications of the general relativistic dispersion relation. Afterwards we will use the perturbative MDRs to calculate specific observables such as the redshift, lateshift and photon circular orbits. 
\end{abstract}


\bodymatter

\section{Traces of quantum gravity in point particle motion}
A large source of information about the gravitational interaction come from the observation of trajectories of freely falling particles, such as for example neutrinos or photons, through spacetime. Traces of the expected quantum nature of gravity and its influence on the particles' paths therefore should manifest itself in the data collected by telescopes and observatories. One approach to derive effects, one can search for in the available data, is provided by quantum gravity phenomenology \cite{AmelinoCamelia:2008qg}. Due to a still missing fundamental theory of quantum gravity, we proceed along the following convincing pictorial idea to set up an effective model for the interaction between point particles and gravity on small distances or high energies: When test particles propagate through spacetime they probe spacetime on lengths scales inverse proportional to their energy. Thus particles with larger energy probe smaller length scales then lower energetic ones. The quantum nature of gravity is expected to become relevant at the Planck scale, i.e.\ at the Planck energy $E_{pl}$ respectively the Planck length $\ell_{pl}$. Hence particles with larger energies, closer to the Planck energy, probe length scales closer to the Planck length and should thus interact stronger with the quantum features of gravity.

This pictorial idea can be seen analogue to how a medium can be probed with photons of different energies, to obtain insights about its constituents. For very low energetic photons the medium may be invisible, while photons within a certain energy range may interact with the medium, get scattered and transport information about the medium to an observer. For the interaction between photons and the elementary constituents of the medium we know, that they are fundamentally explained by the standard model of particle physics. However effectively one can describe several aspects of the system by an energy dependent propagation of the photons through the medium. The latter can be derived from an effective theory of electrodynamics, such premetric electrodynamics \cite{Hehl}, and leads to a non local lorentz invariant (LLI) dispersion relation of the photons. Thus even though a fundamental interaction is LLI, observables may be described effectively by a non LLI theory. 

Regarding the structure of spacetime and the quantum nature of gravity, we do not know the fundamental theory and of quantum gravity and its properties yet. In analogy to the effective description of the propagation of photons through media we will consider modifications of the general relativistic dispersion relation of point particles on spacetime. From such modifications we can calculate observables and look for their signatures in the available data. Any evidence or non-detection of such a signal then reveals properties, suitable semi-classical limits of quantum gravity must have. 

Most famous effects searched for are an energy dependent redshift and time of arrival (lateshift) of photons \cite{Amelino-Camelia:2013uya}. Preliminary analyses of the ICECUBE and Fermi Gamma-Ray Space telescope observations for a lateshift effect have recently been performed \cite{Amelino-Camelia:2016ohi,Xu:2016zsa,Xu:2016zxi,Xu:2018ien} and shall be extended as soon as additional data is available.

We demonstrate how modified dispersion relations (MDRs) are realized on generically curved spacetimes we will derive observables such as the redshift, the lateshift and photon orbits.

\section{Dispersion relations as Hamilton functions}
The study of non LLI effects and MDRs as effective description of the interaction of point particles with the quantum nature of gravity has a long history in the literature, see for example \cite{Matschull:1997du,Freidel:2003sp,Ling:2005bq,Girelli:2006sc,Liberati:2013xla} and references therein. 

In \cite{Barcaroli:2015xda} we demonstrated how to realize modified dispersion relations on curved spacetimes covariantly: \emph{A dispersion relation of a point particle on curved spacetime is given by a level set of a Hamilton function $H(x,p)$ on the point particle phase space (cotangent bundle) of spacetime. The particle's motion is determined by the corresponding Hamilton equations of motion.} Moreover the Hamiltonian defines the geometry of phase space in a canonical way. In general spacetime and momentum space are curved and their curvatures  depend on positions and momenta.

Using this covariant approach to dispersion relations we recall the Hamilton functions of general relativistic particles $H_{GR}$ and construct its first order perturbations $H_{\ell}$ as well as its $\kappa$-Poincar\'e deformations $H_{\kappa}$ \cite{Gubitosi:2013rna,KowalskiGlikman:2001px} on general \cite{Barcaroli:2017gvg}, static spherically symmetric \cite{Barcaroli:2017gvg} and homogeneous and isotropic spacetimes \cite{Barcaroli:2017gvg,Barcaroli:2016yrl,Pfeifer:2018pty}.

On a generically curved spacetime the Hamiltonians are build from a spacetime metric $g$, a perturbation function $h$ and unit timelike vector field of the metric $Z$
\begin{align}
	H_{GR}(x,p) &= g^{ab}(x)p_ap_b\,,\\
	H_{\ell}(x,p) &= g^{ab}(x)p_ap_b + \ell h(x,p) \label{eq:pert1}\,,\\
	H_{\kappa}(x,p) &= -\tfrac{4}{\ell^2}\sinh\big(\tfrac{\ell}{2}Z^c(x)p_c\big)^2 + e^{\ell Z^c(x)p_c} \big(g^{ab}(x)p_ap_b + (Z^d(x)p_d)^2\big)\label{eq:kappa1}\,.
\end{align}
In static spherical symmetry the dependence of the Hamiltonians on the positions and momenta is restricted to $H(x,p) = H(r,p_t,p_r,v)$ with $v^2 =  p_\theta^2 + \sin\theta^{-2} p_\phi^2$ \cite{Barcaroli:2017gvg},
\begin{align}
	H_{GR}(x,p) 
	&= - a(r) p_t^2 + b(r) p_r^2 + r^{-2} v^2\,,\\
	H_{\ell}(x,p)
	&= - a(r) p_t^2 + b(r) p_r^2 + r^{-2} v^2 + \ell h(r,p_t,p_r,v) \label{eq:pert2}\,,\\
	H_{\kappa}(x,p) 
	&=-\tfrac{4}{\ell^2}\sinh\big(\tfrac{\ell}{2}(c(r)p_t + d(r)p_r)\big)^2 + e^{\ell (c(r)p_t + d(r)p_r)}\nonumber \\
	&\times \big( (-a(r) + c(r)^2)p_t^2 + 2 c(r)d(r) p_tp_r + (b(r) + d(r)^2)p_r^2 + \tfrac{1}{r^2}v^2\big)\label{eq:kappa2}\,.
\end{align}
Setting here $a(r)^{-1} = b(r) = 1 - \frac{r_s}{r}$ we obtain the $\kappa$-Poincar\'e deformations of Schwarzschild spacetime parametrized by two free function $c(r)$ and $d(r)$.

\noindent On homogeneous and isotropic spacetimes the form of the Hamiltonian is further restricted to $H(x,p) = H(t,p_t,w)$ with $w^2 =p_r^2 (1 - k r^2) + r^{-2}w^2$
\begin{align}
	H_{GR}(x,p) 
	&=- p_t^2 + A(t)^{-2} w^2\,,\\
	H_{\ell}(x,p)
	&=- p_t^2 + A(t)^{-2} w^2 + \ell h(t,p_t,w) \label{eq:pert3}\,,\\
	H_{\kappa}(x,p) 
	&=-\tfrac{4}{\ell^2}\sinh\big(\tfrac{\ell}{2}p_t\big)^2 + e^{\ell p_t} A(t)^{-2} w^2\label{eq:kappa3}\,.
\end{align}
\section{Observables}
A detection of traces of quantum gravity is most likely with high energetic photons and every MDR of interest can be expanded to first order around $H_{GR}$. Thus we focus on the derivation of observables for massless particles from $H_\ell(x,p)$.

\subsection{Redshift}
The redshift of photons involves a description of how observers measure the frequency of a photon. A classical observer, not subject to the MDR, on a worldline $\xx(\tau)$ with tangent $\dot \xx(\tau)$, momentum $p_\xx(\tau)$ and mass $m_\xx^2 = - H_{GR}(\xx, p_\xx)$, associates to a photon, subject to the MDR, on a worldline $\yy(\tau)$ ,with momentum $p_\yy(\tau)$ and satisfying $H_\ell(\yy,p_\yy) = 0$, the frequency
\begin{align}
	\nu_\xx(\yy) = \dot{\xx}^a p_{\yy a} = \tfrac{1}{2 m_\xx}\partial_{p_a}H_{GR}(\xx, p_\xx)\ p_{\yy a} = \tfrac{1}{m_\xx}g^{ab}(\xx)\ p_{\xx b}\ p_{\yy a}\,.
\end{align}
The redshift of a photon between two observers $\xx_i$ and $\xx_f$ then is 
\begin{align}
	z(t_i,r_i,t_f,r_f) \equiv z =  \tfrac{\nu_{\xx_i}(\yy)}{\nu_{\xx_f}(\yy)} - 1.
\end{align}
Thus, two observers at rest at $\xx_i = (t_i, r_i, \tfrac{\pi}{2},0)$ and $\xx_f = (t_f, r_f, \tfrac{\pi}{2},0)$ find for a radially freely falling photon in spherical symmetry from \eqref{eq:pert2} the redshift \cite{Barcaroli:2017gvg}
\begin{align}
	z =  \sqrt{\tfrac{a(r_f)}{a(r_i)}} - 1 + \mathcal{O}(\ell^2)\,,
\end{align}
while for the homogeneous and isotropic case \eqref{eq:pert3} they find that the redshift becomes dependent of the constant of motion $w$, which is related to the photons frequency \cite{Pfeifer:2018pty}
\begin{align}
	&z = \big(\tfrac{A(t_f)}{A(t_i)}-1\big) - \tfrac{\ell}{2 w^2} \tfrac{A(t_f)}{A(t_i)} \big( A(t_f)^2 h(t_f,p_t^0(t_f,w),w) - A(t_i)^2 h(t_i,p_t^0(t_i,w),w)\big)\nonumber\\
	&\rightarrow \big(\tfrac{A(t_f)}{A(t_i)}-1\big) - \ell w \tfrac{A(t_f) - A(t_i)}{2 A(t_i)^2}\,.
\end{align}
The first order momentum is $p_t^0(t,w) = - \frac{w}{A(t)}$ and the last line displays the result for a first order in $\ell$ expansion of the $\kappa$-Poincar\'e dispersion relation \eqref{eq:kappa3}.

\subsection{Lateshift}
For the so called lateshift consider two massless particles with different frequency parameters $w_1$ and $w_2$ on radial trajectories $r_1(t,w_1)$ and $r_2(t,w_2)$ solving the Hamilton equations of motion of \eqref{eq:pert3}. They shall be emitted at the same coordinates $(t_i,R_i)$, i.e. $R_i = r_1(t_i,w_1)=r_2(t_i,w_2)$, and we ask when do they reach the same radial distance $R_f = r_1(t_1,w_1)=r_2(t_2,w_2)$. The difference in their time of arrival $\Delta t = t_2 - t_1$ is the so called lateshift. For the solutions of the radial Hamilton equation of motion of \eqref{eq:pert3}, derived in \cite{Pfeifer:2018pty}, we find,
\begin{align}\label{eq:lateshift}
\Delta t 
&= \ell A(t_1) \int_{t_i}^{t_1}\mathrm{d}\tau\ \tfrac{f(\tau,p_t^0(\tau,w_2),w_2) - f(\tau,p_t^0(\tau,w_1),w_1)}{A(\tau)}\\
&\rightarrow -\ell \tfrac{A(t_1)(w_1 - w_2)}{2}\int_{t_i}^{t_1}d\tau \tfrac{1}{A(\tau)^2}\,.
\end{align}
where again $p_t^0(t,w) = - \frac{w}{A(t)}$, the last line is the first order in $\ell$ result of \eqref{eq:kappa3}, and
\begin{align}
	f(t,p_t^0(t,w),w) = \tfrac{1}{2 (p_t^0)^2} \big[ h(t,p_t^0,w) - p_t^0 \partial_{p_t} h(t,p_t^0,w) - w \partial_w h(t,p_t^0,w)  \big]\,.
\end{align}

\subsection{Innermost circular photon orbits}
To calculate the innermost circular photon orbits we solve the Hamilton equations of motion of \eqref{eq:pert2} with the assumption that $\dot r = 0$. Due to the spherical symmetry we can without loss of generality consider orbits in the equatorial plane $\theta = \tfrac{\pi}{2}$. The Hamilton equations of motion $\partial_{p_r} H_\ell = \dot r = 0$ and $\partial_r H = - \dot p_r$ with $p_r = p_r^0 + \ell p_r^1$, together with the fact that $p_t$ and $v$ are constants of motion, then imply to first order in $\ell$
\begin{align}
	p_r^0 = 0, \quad p_r^1 = - \tfrac{\partial_{p_r}h(r,p_t,p_r^0,v)}{2b(r)}\quad \Rightarrow\quad \dot p_r^0 = 0 = \dot p_r^1\,.
\end{align}
Using $H_\ell(x,p) = 0$ to express $p_t$ as function of $r$ and $v$, and solving $\partial_r H = - \dot p_r = 0$ yields,
\begin{align}\label{eq:photon}
	0 =\tfrac{a'(r_0)}{a(r_0)} + \tfrac{2}{r_0} ,\quad
	r_1 = \tfrac{\partial_r h(r_0,p^0_t(r_0,v),0,v) -  \tfrac{a'(r_0)}{ a(r_0)} h(r_0,p^0_t(r_0,v),0,v)}{\big(\tfrac{a''(r_0)}{a(r_0)r_0^2} - \tfrac{a'(r_0)^2}{a(r_0)^2 r_0^2} - \tfrac{2 a'(r_0)}{a(r_0)r_0^3} - \tfrac{6}{r_0^4}\big)v^2}  \,,
\end{align}
which determine the allowed circular photon orbits $r = r_0 +\ell r_1$ to first order in $\ell$.

Considering MDRs on Schwarzschild spacetime by specifying the function $a(r)^{-1} = 1 - \tfrac{r_s}{r}$ the zeroth order becomes $r_0 = \frac{3}{2}r_s$, as it must be, and the equation which determines the first order correction becomes
\begin{align}\label{eq:r1}
	r_1 = \tfrac{27 r_s^4(\partial_r h(r_0,p^0_t(r_0,v),0,v) - \tfrac{a'(r_0)}{ a(r_0)} h(r_0,p^0_t(r,v),0,v))}{32 v^2}\,.
\end{align}
Specifying to the first order in $\ell$ of the spherically symmetric $\kappa$-Poincar\'e dispersion relation \eqref{eq:kappa2} with $d(r) = 0$ and $c(r) = \frac{1}{\sqrt{1-\frac{r_s}{r}}}$ \eqref{eq:r1} yields $r_1 =\frac{v}{6}$ \cite{Barcaroli:2017gvg}.

\subsection{Conclusion}
The implementation of modified dispersion relation on curved spacetimes allows for the prediction of traces of the quantum nature of gravity, which can be confirmed or constraint by observations. Particularly promising are time of arrival measurements of high energetic gamma-rays, which can be compared against equation \eqref{eq:lateshift} and the search for rainbow effects in lensing images, for example in the shadows of black holes, which would correspond to a non-trivial $v$ dependence in \eqref{eq:photon}. 

To bridge the the gap between fundamental approaches to quantum gravity and the phenomenological one presented here, it is necessary to derive the perturbation function $h(x,p)$ in \eqref{eq:pert1} in some semi-classical limit from the fundamental approaches. This is one of the major tasks for future research.

\section*{Acknowledgments}
This talk is based on a series of articles \cite{Barcaroli:2015xda,Barcaroli:2016yrl,Barcaroli:2017gvg} written in collaboration with Leonardo Barcaroli, Lukas Brunkhorst, Giulia Gubitosi and Niccol\'o Loret.


\begin{thebibliography}{0}

\bibitem{AmelinoCamelia:2008qg}
G.~Amelino-Camelia,
Living Rev.\ Rel.\  {\bf 16} (2013) 5
doi:10.12942/lrr-2013-5
[arXiv:0806.0339 [gr-qc]].

\bibitem{Hehl}
F. W. Hehl and Y. N. Obukhov,
``Foundations of Classical Electrodynamics,''\\
Birkh\"auser, 2003.

\bibitem{Amelino-Camelia:2013uya}
G.~Amelino-Camelia, L.~Barcaroli, G.~Gubitosi and N.~Loret,
Class.\ Quant.\ Grav.\  {\bf 30} (2013) 235002
doi:10.1088/0264-9381/30/23/235002
[arXiv:1305.5062 [gr-qc]].


\bibitem{Amelino-Camelia:2016ohi}
G.~Amelino-Camelia, G.~D'Amico, G.~Rosati and N.~Loret,
Nat.\ Astron.\  {\bf 1} (2017) 0139
doi:10.1038/s41550-017-0139
[arXiv:1612.02765 [astro-ph.HE]].

\bibitem{Xu:2018ien} 
H.~Xu and B.~Q.~Ma,
JCAP {\bf 1801}, no. 01, 050 (2018)
doi:10.1088/1475-7516/2018/01/050
[arXiv:1801.08084 [gr-qc]].

\bibitem{Xu:2016zsa}
H.~Xu and B.~Q.~Ma,
Phys.\ Lett.\ B {\bf 760} (2016) 602
doi:10.1016/j.physletb.2016.07.044
[arXiv:1607.08043 [hep-ph]].

\bibitem{Xu:2016zxi}
H.~Xu and B.~Q.~Ma,
Astropart.\ Phys.\  {\bf 82} (2016) 72
doi:10.1016/j.astropartphys.2016.05.008
[arXiv:1607.03203 [hep-ph]].

\bibitem{Matschull:1997du}
H.~J.~Matschull and M.~Welling,
Class.\ Quant.\ Grav.\  {\bf 15} (1998) 2981
doi:10.1088/0264-9381/15/10/008
[gr-qc/9708054].

\bibitem{Freidel:2003sp}
L.~Freidel, J.~Kowalski-Glikman and L.~Smolin,
Phys.\ Rev.\ D {\bf 69} (2004) 044001
doi:10.1103/PhysRevD.69.044001
[hep-th/0307085].

\bibitem{Ling:2005bq}
Y.~Ling, B.~Hu and X.~Li,
Phys.\ Rev.\ D {\bf 73} (2006) 087702
doi:10.1103/PhysRevD.73.087702
[gr-qc/0512083].

\bibitem{Girelli:2006sc}
F.~Girelli, S.~Liberati, R.~Percacci and C.~Rahmede,
Class.\ Quant.\ Grav.\  {\bf 24} (2007) 3995
doi:10.1088/0264-9381/24/16/003
[gr-qc/0607030].

\bibitem{Liberati:2013xla}
S.~Liberati,
Class.\ Quant.\ Grav.\  {\bf 30} (2013) 133001
doi:10.1088/0264-9381/30/13/133001
[arXiv:1304.5795 [gr-qc]].

\bibitem{Barcaroli:2015xda}
L.~Barcaroli, L.~K.~Brunkhorst, G.~Gubitosi, N.~Loret and C.~Pfeifer,
Phys.\ Rev.\ D {\bf 92} (2015) no.8,  084053
doi:10.1103/PhysRevD.92.084053
[arXiv:1507.00922 [gr-qc]].

\bibitem{Gubitosi:2013rna} 
G.~Gubitosi and F.~Mercati,
Class.\ Quant.\ Grav.\  {\bf 30}, 145002 (2013)
doi:10.1088/0264-9381/30/14/145002
[arXiv:1106.5710 [gr-qc]].

\bibitem{KowalskiGlikman:2001px}
J.~Kowalski-Glikman,
Mod.\ Phys.\ Lett.\ A {\bf 17} (2002) 1
doi:10.1142/S0217732302006175
[hep-th/0107054].

\bibitem{Barcaroli:2017gvg}
L.~Barcaroli, L.~K.~Brunkhorst, G.~Gubitosi, N.~Loret and C.~Pfeifer,
Phys.\ Rev.\ D {\bf 96} (2017) no.8,  084010
doi:10.1103/PhysRevD.96.084010
[arXiv:1703.02058 [gr-qc]].

\bibitem{Barcaroli:2016yrl}
L.~Barcaroli, L.~K.~Brunkhorst, G.~Gubitosi, N.~Loret and C.~Pfeifer,
Phys.\ Rev.\ D {\bf 95} (2017) no.2,  024036
doi:10.1103/PhysRevD.95.024036
[arXiv:1612.01390 [gr-qc]].

\bibitem{Pfeifer:2018pty}
C.~Pfeifer,
Phys.\ Lett.\ B {\bf 780} (2018) 246
doi:10.1016/j.physletb.2018.03.017
[arXiv:1802.00058 [gr-qc]].

\end{thebibliography}
\end{document}